\documentclass[prl,twocolumn,showpacs,preprintnumbers,amsmath,amssymb]{revtex4}
\usepackage{graphicx}
\usepackage{latexsym}
\usepackage{color}
 
\begin{document}
\title{Measurement of particle and bubble accelerations in turbulence}

\author{R. Volk$^{(1)}$, N. Mordant$^{(2)}$, G. Verhille$^{(1)}$ and J.-F. Pinton$^{(1)}$}
\affiliation{$^1$ Laboratoire de Physique, de l'\'Ecole normale
sup\'erieure de Lyon, CNRS UMR5672, 46 All\'ee d'Italie, 69007 Lyon, France\\
$^2$ Laboratoire de Physique Statistique de l'\'Ecole normale sup\'erieure de Paris, CNRS UMR8550, 
24 rue Lhomond, 75005 Paris, France}

\begin{abstract}
We use an extended laser Doppler technique to track optically the velocity of individual particles in a high Reynolds number turbulent flow. The particle sizes are of the order of the Kolmogorov scale and the time resolution, 30 microseconds, resolves the fastest scales of the fluid motion. Particles are tracked for mean durations of the order of 10 Kolmogorov time scales. The fastest scales of the particle motion are resolved and the particle acceleration is measured. For neutrally buoyant  particles,  our measurement matches the performance of the silicon strip detector technique introduced at Cornell University~\cite{Voth,MordantCornell}.  This reference dynamics is then compared to that of slightly heavier solid particles (density 1.4) and to air bubbles. We observe that the acceleration variance strongly depends on the particle density: bubbles experience higher accelerations than fluid particles, while heavier particles have lower accelerations. We find that the probability distribution functions of accelerations normalized to the variance are very close although the air bubbles have a much faster dynamics.
\end{abstract}

\maketitle

The Lagrangian approach to fluid dynamics is a natural one when one addresses problems related to mixing and transport~\cite{lagmix}. It has also been widely studied in the context of intermittency in fully developed turbulence. In recent years, several novel experimental techniques have been developed. The pioneering optical tracking method developed in the Cornell group has revealed that fluid particles experience extremely intense accelerations, with probability density functions (PDFs) having stretched exponential tails~\cite{Voth,MordantCornell}. Initially limited to very short particle tracks, the technique has been extended with the use of ultrafast optical cameras~\cite{FastCameras}, and is currently applied to the study of multiple particle statistics~\cite{Multipart}.  Individual particles have been tracked for time duration of the order of the flow integral time scale using an acoustic technique~\cite{lyon1}: in an insonified volume, individual particles scatter a sound wave whose Doppler shift carries the tracer velocity. In reason of the very fast decrease of the acoustic scattering cross-section, this method is limited to particles with size of the order of the wavelength, {\it i.e.}  inertial ranges sizes~\cite{lyon2} when using acoustics. However the principle of the technique is completely analogous to laser Doppler velocimetry (LDV), provided that expanded light beams are used. Interference fringes are created at the intersection of two wide laser beams; a particle that crosses these fringes scatters light with a modulation frequency proportional to its velocity component perpendicular to the direction of the fringes~\cite{Tropea}. The advantage, compared to the acoustic method, is that the much smaller wavelength allows a better resolution in space and also the use of smaller tracer particles. 

In this Letter, we describe the principles of this technique and validate it against the known features of the Lagrangian acceleration statistics in a fully turbulent von K\'arm\'an flow  at $R_\lambda \leq 850$. We then apply it to track the dynamics of particles whose density differs from that of the fluid. The dynamics of such {\it inertial} particles is relevant for many engineering applications related to transport, mixing, dispersion, etc~\cite{EngInertial}. Significant theoretical  and numerical progress in this domain has been made in the limiting case of infinitely heavy, pointwise particles~\cite{TheoInertial} and has received  experimental verifications~\cite{Eaton,Warhaft}. 
We report the first experimental measurements of accelerations of particles having a density in the range $10^{-3}$ (air bubbles) to $1.4 $ (PMMA) in the same highly turbulent flow. Taking into account the added mass effect for small spherical particles (i.e. the displacement of fluid elements by the particle motion), the effective density of the bubbles is only 3 times less than that of the fluid, while the PMMA particles are roughly 1.5 times larger. Because of the mismatch of density, light particle tend to be trapped in high vortical regions:  as  result of a lower inertia, the centrifugal force can not compensate the pressure gradient which drive them into the core of the vortices. On the contrary the centrifugal force is stronger than the pressure gradient for heavy particles so that they are ejected form vortex cores and concentrate in high strain regions~\cite{SundaramCollins}. Because of this distinct spatial sampling of the flow, particle with different buoyancy are expected to exhibit quite different dynamical behavior. Indeed, we do find that the particles have different dynamical characteristics such as acceleration variance or correlation time. The PDFs of their accelerations remain close for value less than about 10 times the acceleration variance, and differ for higher values.

\begin{figure}
\includegraphics[width=7.5cm]{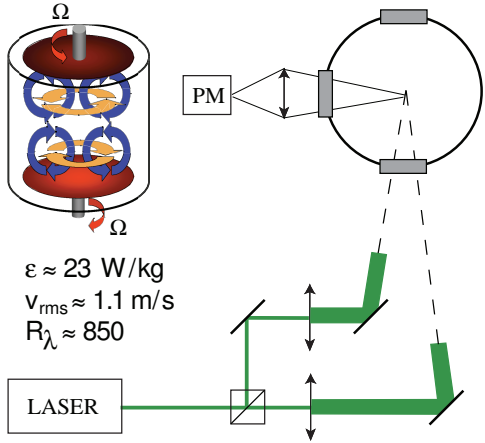}
\caption{Experimental setup. (left): schematics of the von K\'arm\'an flow in water -- side view. (right): principle of the Laser-Doppler Velocimetry using wide beams (ELDV) -- top view of the experiment. PM: location of the photmultipler which detect scattering light modulation as a particle crosses the interference pattern created at the intersection of the laser beams.}
\end{figure}

The Laser Doppler technique is based on the same principle as the ultrasound Doppler method which was shown to be very valuable for Lagrangian particle tracking~\cite{lyon1}. 
I order to access dissipative scales, and in particular for acceleration measurements, one adapts the technique from ultrasound to Laser: the gain is of  a factor 1000 in wavelength so that one expects to detect micron-sized particles. For a Lagrangian measurement, one has to be able to follow the particle motion to get information about its dynamics in time. For this, wide Laser beams are needed to illuminate the particle on a significant fraction of its path. The optical setup is an extension of the well known laser Doppler velocimetry technique; cf. Fig.1. A Laser beam is split into two beams; each is  then expanded by a telescope so that their diameter is about 5mm. Then the two beams are directed in the flow. In their intersection volume, they create an array of interference fringes. As a particle goes across the fringes, the scattered light is modulated and the frequency modulation is directly proportional to the component of the velocity perpendicular to the interference fringes. One then measures one component of the particle velocity. 
In practice, we use a CW YAG laser of wave length 532 nm with 1.2 W maximum output  power. In order to get the sign of the velocity, the standard method consists in using a acousto-optic modulator (AOM) to shift the frequency of one of the beams so that the fringes are actually travelling at a constant speed. Here we use one AOM for each beam, the two excitation frequencies of the AOM being shifted by 100~kHz. The angle of the two beams is tuned to impose a 60 microns interfringe so that 100kHz corresponds to 6~m/s. As the beams are not focused, the interfringe remains constant across the measurement volume whose size is about $5 \times 5 \times 10 \textrm{ mm}^3$. The measurement volume is imaged on a photomultiplier whose output is recorded using a National Instrument PXI-NI5621 digitizer.

The flow is of the Von K\'arm\'an kind as for the ultrasound measurements~\cite{lyon1}.  Water fills a cylindrical container of internal diameter 15~cm, length~20 cm. It is driven by two disks of diameter 10 cm, fitted with blades. The rotation rate is fixed at values up to 10~Hz. For the measurements reported here, the Taylor based Reynolds number up to 850 and the dissipation rate $\epsilon$ up to 25 W/kg. We study three types of particles: neutrally buoyant polystyrene particles of size 31 microns and density 1.06, PMMA particles of size 43 microns and density 1.4 and air bubbles with a size of about 100 microns. The size of the bubbles, measured optically by imaging the measurement volume on a CCD, is imposed by the atomisation of a large bubble by the turbulent flow, a process known to lead to a well defined and stationary size distribution.

\begin{figure}
\includegraphics[width=8.5cm]{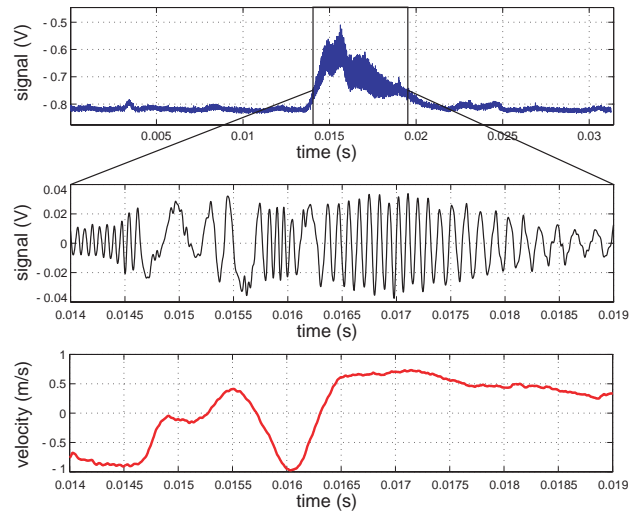}
\caption{Signal processing. (top): raw light modulation, as detected by the photodiode when a particle crosses the fringes. (middle): heterodyne detection of the frequency modulation. (bottom): velocity trace extracted using the AML algorithm~\cite{JASA}. }
\end{figure}

The signal processing step is crucial as both time and frequency -- ({\it i.e.}  velocity -- resolutions rely on its performance. Frequency demodulation is achieved using the same algorithm as in the acoustic Doppler technique. It is a approximated maximum likelihood method coupled which a Kalman filter~\cite{JASA}: a parametric estimator assumes that the signal is made of a modulated complex exponential and Gaussian noise. The amplitude of the sound and the modulation frequencies are assumed to be slowly evolving compared to the duration of the time window used to estimate the instantaneous frequency. Here the time window is about 30~$\mu$s long and sets the time resolution of the algorithm. Outputs of the algorithm are the instantaneous frequency, the amplitude of the modulation and a confidence estimate which can be used to discriminate bad data. An example of the light scattered by a particle is displayed in Fig.~2, together with the Doppler frequency modulation, and final velocity signal. Afterwards, the acceleration of the particle is computed by differentiation of the velocity output, using a low pass filter to smooth the noise as in~\cite{MordantCornell}. After processing, the data consists in a collection of sequences whose lengths are exponentially distributed. 

\begin{figure}
\centerline{\includegraphics[width=7cm]{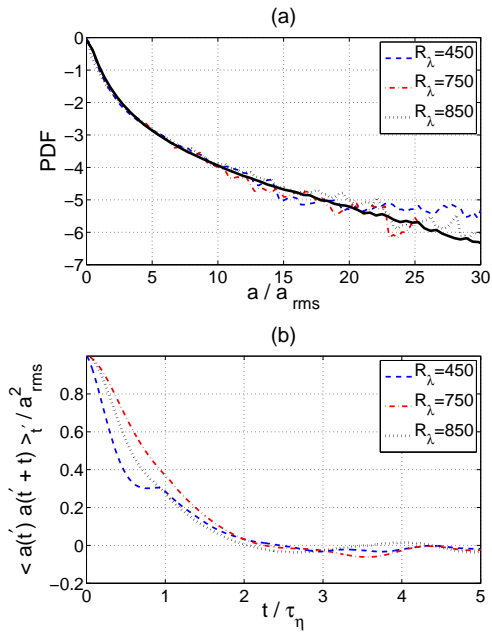}}
\caption{(a) Probability density functions (PDFs) of the acceleration, normalized by its variance,  for neutral particles. (symbols): ELDV measurements; (black thick line): Cornell data at $R_\lambda = 690$.   (b) autocorrelation coefficients of the accelerations for the same cases.}
\end{figure}

We first compare the data for neutral particles to the existing data recorded in a similar flow in Cornell University using high speed imaging~\cite{Voth,MordantCornell}: the measurement used linear cameras of 512 pixels running at speeds up to $70,000$ frames per seconds to get the time and linear resolution required to resolve the acceleration. The probability density functions (PDFs) of the acceleration, from our ELDV technique at increasing Reynolds numbers and for  the Cornell data at $R_\lambda = 690$ are displayed in Fig.~3. The distributions are seen to be very good agreement both qualitatively and quantitatively. In order to compare the two experiments, the following Heisenberg-Yaglom scaling is used \begin{equation}
\langle a^2 \rangle = a_0 \epsilon^{3/2}\nu^{-1/2} \ , 
\end{equation} 
where $\epsilon$ is the energy dissipation rate per unit mass and $\nu = 1.3 \cdot 10^{-6}$~m$^2$.s$^{-1}$ is the kinematic viscosity of the fluid. We derive here $a_0 = 6.4 \pm 1$ at $R_\lambda=850$ compared to $6.2\pm 0.4$ for the Cornell data at $R_\lambda = 690$. The acceleration variance is computed using the same procedure as in~\cite{MordantCornell}: it is obtained for several width of the smoothing kernel used in the differentiation of the velocity signal and then interpolated to zero filter width. The acceleration autocorrelation function, shown in Fig.3b, decays in a time of the order of the Kolmogorov time $\tau_\eta=\sqrt{\nu/\epsilon}$, the fastest timescale of the turbulent flow -- for the data shown here, $\tau_\eta=0.25$ms. This confirms that our techniques achieves a fast enough time and velocity resolution to get a good estimate of the acceleration.  

\begin{table}
\begin{center}
\begin{tabular}{|c c c c c c|}
\hline  $\Omega$ & $u_{\rm rms}$ & $a_{\rm rms}$ & $\epsilon$ &  $R_\lambda$ & $a_0$\\
\hline  $[{\rm Hz}]$ & $[{\rm m.s}^{-1}]$ & $[{\rm m.s}^{-2}]$ & $[{\rm W.kg}^{-1}]$ &  - & - \\
\hline  4.1 & 0.5 & 227 &  4  & 450 & $4 \pm 1.5$ \\
\hline  6.4 & 0.8 & 352 &10  & 750 & $4.2 \pm 1$ \\
\hline  8.9 & 1.1 & 826 & 23  & 850  & $6.4 \pm 1$ \\
\hline
\end{tabular}
\end{center}
\caption{Parameters of the flow. $\Omega$: rotation rate of the disks, $\epsilon$  dissipation rate obtained from the power consumption of the motors (with an accuracy of about $20\%$). The Taylor-based turbulent Reynolds number is computed as $R_\lambda=\sqrt{{15 u_{rms}^4}/{\epsilon\nu}}$, and $a_0$ is derived from the Heisenberg-Yaglom relationship -- Eqn.~1.  }
\end{table}

The very close agreement between the measurements reported here and the Cornell~\cite{MordantCornell} data is expected because they share the same flow geometry. The fact that it is observed using two very different techniques and signal processing validates both (as the performance of the high speed imaging method had not been matched previously). 

\begin{figure}
\centerline{\includegraphics[width=7.5cm]{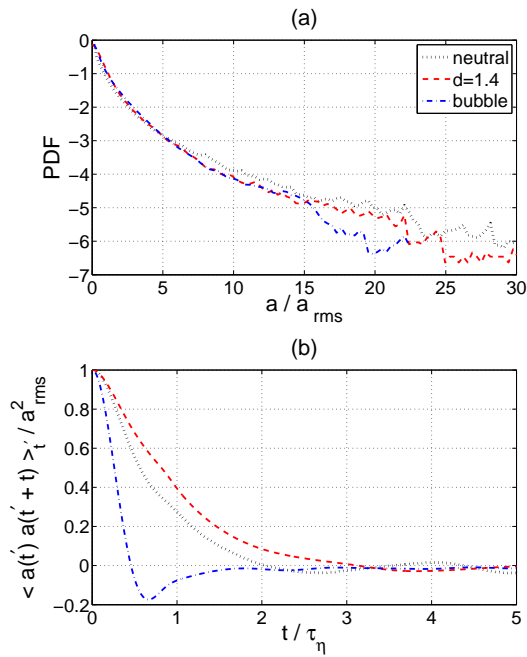}}
\caption{(a) Normalized autocorrelation coefficient of the acceleration. (b) Probability distribution function of accelerations, normalized to the variance of the data sets. Flow at $R_\lambda = 850$.}
\end{figure}

We now apply our technique to compare Lagrangian tracers to the dynamics of heavier and lighter particles. We first compute the velocity root mean square value $u_{rms}$ for all three cases: the values are $1.1, 1.2, 1.0 \pm 0.1 \, {\rm  m.s}^{-1}$ at $R_\lambda=850$ for the neutral, heavy (PMMA spheres) and light particles (bubbles). Within error bars, the large scale dynamics seems to be  unaffected by changes in the particle inertia. The acceleration distribution and autocorrelation in the three cases is shown in Fig.4. The acceleration PDFs are quite similar for moderate acceleration values (below about $10 a_{\rm rms}$), as also observed in low Reynolds number numerical simulations~\cite{Lohse}. However, the probability of very large accelerations is reduced in the case of bubbles compared to neutral particles. The normalized acceleration variance $a_0$ varies very significantly: it is reduced  to $4.3 \pm 1$  for heavier particles at $R_\lambda=850$ while it is increased  to $26 \pm 5$ for bubbles. The correlation functions also show significant changes with the inertia: the characteristic time of decay is longer for heavy particles and shorter for bubbles compared to neutral particles. We measure $\tau_{\rm corr}/\tau_\eta =  0.5, \; 0.9, \; 0.25$ respectively for neutral, heavy and light particles, with the correlation time defined as the half-width at mid amplitude of the correlation function. We thus observe significant changes in the dynamics, even if the distribution of acceleration weakly changes with inertia. 

We now briefly discuss these results.  In order to characterize the dynamics of a solid particle, one must specify two dimensionless numbers: the Stokes number which, in the case of turbulent flows, is the ratio of the response time of the particle to the Kolmogorov time scale, and the density ratio of the particle's material to that of the fluid. In the asymptotic case of very large inertia, only the Stokes number matters~\cite{TheoInertial}. Here the inertia remains finite even for the bubble case because of the added mass contribution. In the limit of very small particles compared to the Kolmogorov length scale one can derive an effective equation of motion for a solid particle~\cite{Maxey}:
\begin{equation}
\frac{d\mathbf v_p}{dt}=\beta \frac{D\mathbf u}{Dt}+\mathbf F \ ,
\label{eqv}
\end{equation}
where $v_p$ is the particle velocity, $\beta={3\rho_f}/({2\rho_p+\rho_f})$, $\rho_f$ and $\rho_p$ are the fluid and particle specific mass respectively), ${D\mathbf u}/{Dt}$ is the acceleration of the fluid particle that would be at the position of the solid particle in the undisturbed flow and $\mathbf F$ incorporates other forces such as drag, lift, history and possibly the buoyancy (negligible here). We have $\beta=1$, 0.8 and 3 for the neutral, heavy and light particles. Bubble with such a small diameter are usually considered as being rigid because of impurities in the fluid~\cite{Duineveld} but the boundary condition may be different from that of a solid particle. We find that the trend for the acceleration variance follows qualitatively that of $\beta$ in eqn~(\ref{eqv}). Quantitatively, for heavy particles it changes roughly as $\beta$, but for bubbles, it is only about 2.6 times that of the fluid. This is different from what was reported in~\cite{vothjfm} where a stronger influence of small change in the particles density was observed.

To conclude, we have reported here an extended Laser Doppler velocimetry technique (ELDV) for the tracking of individual particles in fully developed turbulence. The advantage of our technique is that it may be more easily adapted from commercial equipment than ultrafast PIV, and requires less laser power for illumination. Its application to the study  of inertial particles with a finite density with respect to the fluid shows that quite different dynamics may lead to very similar statistics of particle accelerations, an observation that may prove useful for modeling.

{\bf Acknowledgement} We are indebted to Artem Petrosian for his help in setting-up up the optics of the experiment, and to Mickael Bourgoin for many fruitful discussions. This work was partially funded by the Region Rh\^one-Alpes, under Emergence Contract No. 0501551301.

\end{document}